\documentclass[aps,prl,twocolumn,showpacs,superscriptaddress,floatfix]{revtex4}
\usepackage{color,graphicx}
\begin{document}

  \author{S.\ Miyahara}
  \affiliation{
    Multiferroics Project (MF), ERATO,
    Japan Science and Technology Agency (JST), 
    c/o Department of Applied Physics, The University of Tokyo, 
    7-3-1 Hongo, Tokyo 113-8656, Japan}
  \author{N.\ Furukawa}
  \affiliation{
    Multiferroics Project (MF), ERATO,
    Japan Science and Technology Agency (JST), 
    c/o Department of Applied Physics, The University of Tokyo, 
    7-3-1 Hongo, Tokyo 113-8656, Japan}
  \affiliation{
    Department of Physics and Mathematics, 
    Aoyama Gakuin University, 
    5-10-1 Fuchinobe, Chuo-ku, Sagamihara, Kanagawa 252-5258, Japan}

  \title{
    Theory of magnetoelectric resonance in two-dimensional  $S=3/2$  
    antiferromagnet ${\rm Ba_2CoGe_2O_7}$ 
    via spin-dependent metal-ligand hybridization mechanism
    }
    %spectra 
  \date{\today}

  \begin{abstract}
    We investigate magnetic excitations 
    in an $S=3/2$ Heisenberg model representing two-dimensional 
    antiferromagnet ${\rm Ba_2CoGe_2O_7}$.
    In terahertz absorption experiment of the compound,
    Goldstone mode as well as
    novel magnetic excitations,
    conventional magnetic resonance at 2~meV and
    both electric- and magnetic-active excitation at 4~meV, 
    have been observed. 
    By introducing a hard uniaxial anisotropy 
    term $\Lambda (S^z)^2$, three modes can be explained naturally.
    We also indicate that, via
    the spin-dependent metal-ligand hybridization mechanism,
    the 4~meV excitation is an electric-active
    mode through the coupling between spin and electric-dipole. 
    Moreover, at 4 meV excitation, 
    an interference between magnetic and electric responses
    emerges as a cross correlated effect.
    Such cross correlation effects explain the
    non-reciprocal linear directional dichroism observed in ${\rm Ba_2CoGe_2O_7}$.
  \end{abstract}


  \maketitle 

  Multiferroic materials
  have attracted both experimental and 
  theoretical interests due to giant magnetoelectric 
  effects~\cite{tokura06,eerenstein06,cheong07}.
  Such strong couplings between magnetism and
  electric polarization (EP) are often realized through 
  spin-dependent EPs.
  For example, in cycloidal magnets $R$MnO$_3$ ($R = $ Tb, Dy, and others),
  EP flops from $P \| c$ 
  to $P \| a$ by changing a magnetic state from $bc$ to $ab$ cycloidal 
  state through external magnetic fields~\cite{kimura03}.
  Another example is magnetic resonance induced by
  oscillating electric field, or {\em electromagnon},
  which is observed in an optical spectroscopy at terahertz (THz) frequencies
  for a variety of multiferroics compounds, 
  e.g., $R$MnO$_3$~\cite{pimenov06,kida09,aguilar09}
  and ${\rm CuFe_{1-x}Ga_xO_2}$~\cite{seki10}. 
  The exchange striction~\cite{aguilar09,arima06}
  and the spin current~\cite{katsura05,mostovoy06,sergienko06} mechanisms
  are well known as the origins of
  such spin-dependent EP.

  Spin-dependent metal-ligand hybridization 
  has been proposed as an alternative mechanism~\cite{jia06,arima07}.
  EP along the bond direction ${\bf r}_{ml}$ connecting metal and ligand 
  depends on a spin structure at a metal site ${\bf S}_m$ 
  in a form
  ${\bf p}_{ml} \propto ({\bf S}_m \cdot {\bf r}_{ml})^2 {\bf r}_{ml}$.
  At a spin site with no inversion,
  such a mechanism can induce an electric dipole which
  is coupled to the spin,
  and has a potential to induce novel features.
  In fact, magnetic field dependence of 
  the ferroelectricity 
  observed in ${\rm Ba_2CoGe_2O_7}$ can well be explained 
  by introducing this mechanism~\cite{murakawa10}. 

\begin{figure}
    \begin{center}
      \includegraphics[width=0.98\columnwidth]{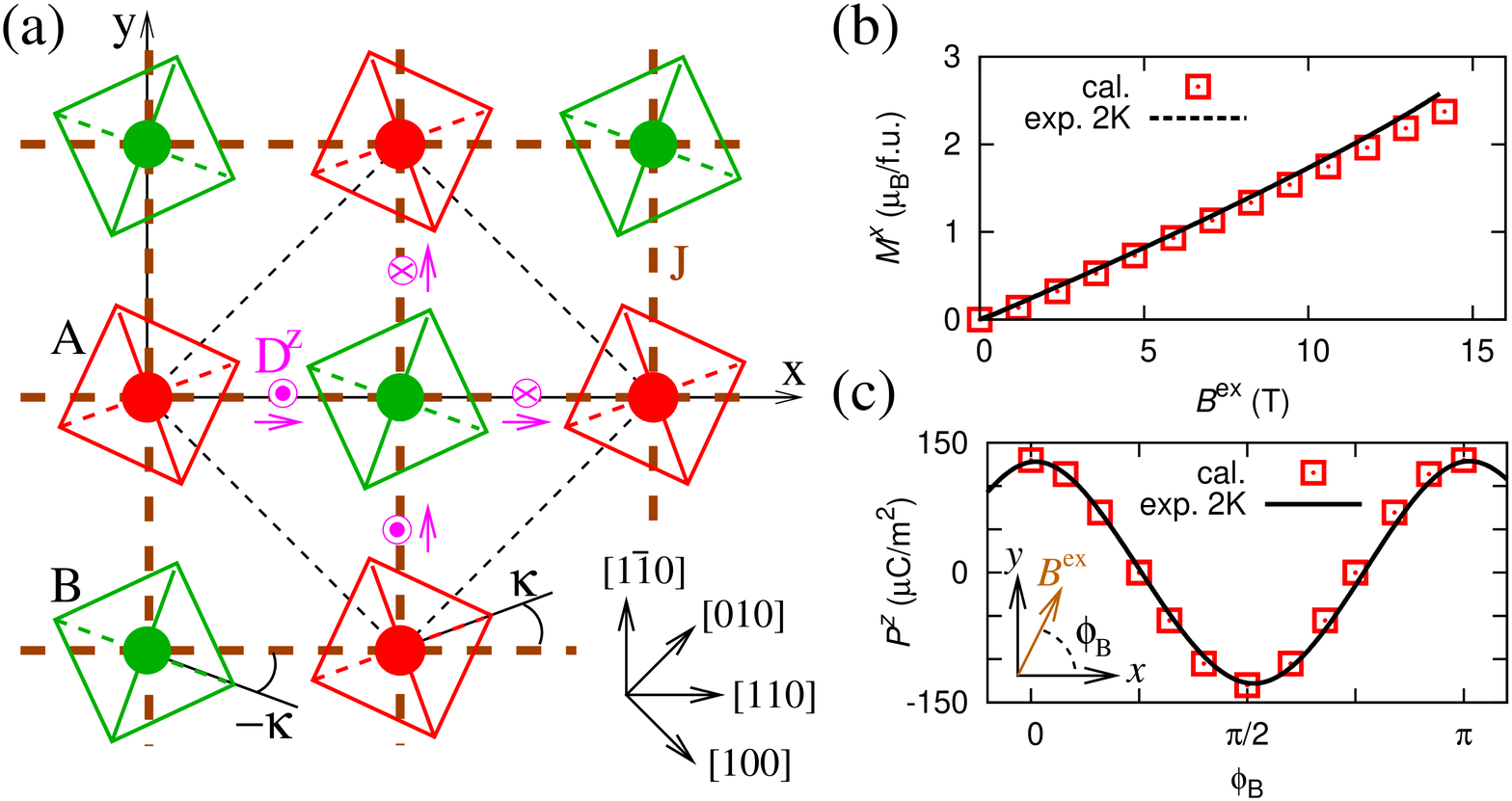}
    \end{center}
    \caption{(Color online) 
      (a) Crystal structure for ${\rm Ba_2CoGe_2O_7}$. 
      Top view of ${\rm CoO_4}$ tetrahedra is illustrated 
      by squares.
      An $S = 3/2$ spin locates on a Co-site with a spin interaction $J$.
      DM interaction $- D^z_{ij} \, (S^x_i S^y_{j} - S^y_i S^x_{j})$
      is also included.
      On each bond, positive (negative) sign of $D^z_{ij}$ 
      is represented by $\odot$ ($\otimes$), and
      the direction from $i$-  to $j$-site  is indicated by the arrow.
      The tetrahedra ${\rm CoO_4}$ on $A$- and $B$-sublattices
      are rotated around $z$ axis with a rotation angle 
      $\kappa$ and $-\kappa$, respectively. 
      See also Ref.~\cite{murakawa10}.
      (b) Magnetization curve for ${\bf B}^{\rm ex} \| x$.
      (c) EP along $z$-direction
      in the in-plane magnetic field ${\bf B}^{\rm ex} 
      = (B^{\rm ex} \cos \phi_{\rm B}, B^{\rm ex} \sin \phi_{\rm B}, 0)$ 
      ($B^{\rm ex} = 5$ T) on a 12-site cluster.
      Experimental data in (b) and (c) are 
      extracted from Ref.~\onlinecite{murakawa10} and
      Ref.~\onlinecite{murakawa10b}, respectively.
    }
    \label{fig:model}
  \end{figure}

  ${\rm Ba_2CoGe_2O_7}$ is a quasi two-dimensional antiferromagnet
  (Fig.~\ref{fig:model} (a)).
  Below $T_N = 6.7$ K, Co magnetic moments ($S = 3/2$) show 
  an antiferromagnetic structure, 
  where magnetic moments are aligned in $xy$-plane
  due to an easy-plane anisotropy~\cite{zheludev03}.
  In the magnetically ordered state, peculiar 
  magnetoelectric behaviors have been observed~\cite{murakawa10,yi08}.
  For example,  EP along $[001]$ shows sinusoidal 
  angular dependence
  with a period of $\pi$ for a rotation of the magnetic 
  field ${\bf B}^{\rm ex}$ within $xy$-plane at $B^{\rm ex} \gtrsim 1$ T.
  As shown in Ref.~\onlinecite{murakawa10},  
  such magnetoelectric behaviors can well be explained
  by a local electric dipole moment which couples 
  to the local spin structure of Co atom
  via the metal-ligand hybridization mechanism between Co and O atoms.
  %$p_m^{x'} = -K ( S_m^{z'} S_m^{x'} + S_m^{x'} S_m^{z'} )$, 
  %$p_m^{y'} = K ( S_m^{y'} S_m^{z'} + S_m^{z'} S_m^{y'} )$, 
  %and $p_m^{z'} = K \{ (S_m^{y'})^2 - (S_m^{x'})^2 \}$
  %where coordinates $x', y'$ and $z'$ are defined as in
  %Fig.~\ref{fig:model} (b) for a ${\rm CoO_4}$ tetrahedron.
  For a classical spin within the $xy$-plane 
  ${\bf S}_m = ( S \cos \theta, S \sin \theta, 0)$, 
  an EP on a ${\rm CoO_4}$ tetrahedron
  along $z$ is described as $p_m^{z} = -S^2 K \cos \, 2 \theta$,
  which reproduces the experimental results.
%  Such a feature is peculiar for
%  the spin-dependent metal-ligand hybridization mechanism.
  
  However, there still exist several features to be understood.
  One of them is magnetic excitation property observed in 
  an electromagnetic wave (EMW) absorption 
  experiment (AE) in the THz frequency regime,
  which indicates magnetic resonances at
  $\omega \sim 0$~\cite{kezsmarki11}, $2$, and 4 meV~\cite{kezsmarki10}. 
  The lowest two peaks can be assigned to spin wave branches
  which have been reported in the inelastic neutron 
  scattering experiments (INS)~\cite{zheludev03}. 
  Here, two distinct modes exist at the $\Gamma$ point 
  in the two-sublattice ground state
  due to an anisotropy.
  However, the origin of the excitation at $\omega \sim 4$ meV
  is not clear within magnon pictures.
  The other point to be understood is 
  the THz AE on several EMW 
  polarizations which indicates that the excitation 
  at $4$ meV is induced by both magnetic and
  electric components of EMW, whereas 
  the excitation at $2$ meV is excited mainly 
  by the in-plane magnetic component.
  Moreover, the resonance at $4$ meV
  shows a non-reciprocal directional dichroism (NDD)
  under the external magnetic fields~\cite{kezsmarki10},
  i.e., absorption intensity strongly depends on the EMW
  propagation directions (forward $+{\bf k}$ or backward $-{\bf k}$).
  In contrast, NDD is not clearly observed at the 2~meV resonance.
  The origin of the magnetic excitation and 
  the absorption mechanism is very important to understand 
  the principle of the NDD.

  In this Letter, we propose that 
  a uniaxial anisotropy term $\Lambda (S_m^z)^2$ ($\Lambda > 0$)
  gives clues to understand these features.
  In an $S = 3/2$ system,  the uniaxial anisotropy splits single spin energies 
  into two doubly degenerate states:
  $|\pm \frac{1}{2} \rangle$ with an eigenenergy
  $\Lambda/4$ and $|\pm \frac{3}{2} \rangle$ with $9\Lambda/4$.
  Here, $|m\rangle$ is a state with $S_z=m$ for spin $S=3/2$.
  In the strong anisotropy limit $J/\Lambda \ll 1$,
  where we neglect the higher energy spin states
  $|\pm \frac{3}{2} \rangle$,
  an $S=3/2$ Heisenberg model
  can be approximated by an XXZ model ${\cal H}^{\rm eff} = \sum J \{
  \Delta (\sigma_i^x \sigma_j^x + \sigma_i^y \sigma_j^y)
  + \sigma_i^z \sigma_j^z \}$  with $\Delta = 4$ by using 
  an $S=1/2$ spin operator ${\bf \sigma}$~\cite{memo-Pz}.
  It should be a good approximation
  to reproduce the lowest two branches of excitations.
  In fact, $\Delta \sim 2.5$ gives a good fit to neutron data
  in Ref.~\onlinecite{zheludev03}.
%  Due to the anisotropy, 
%  a finite energy excitation exists around 2 meV at the $\Gamma$ point,
%  whereas the Goldstone mode remains nearly gapless.
%  In the Heisenberg model with $\Lambda/J \gg 1$,
  On the other hand, 
  the highest energy mode at 4 meV can be assigned to
  magnetic excitation due to the single ion 
  anisotropy gap $2 \Lambda$.

\begin{figure}
    \begin{center}
      \includegraphics[width=0.98\columnwidth]{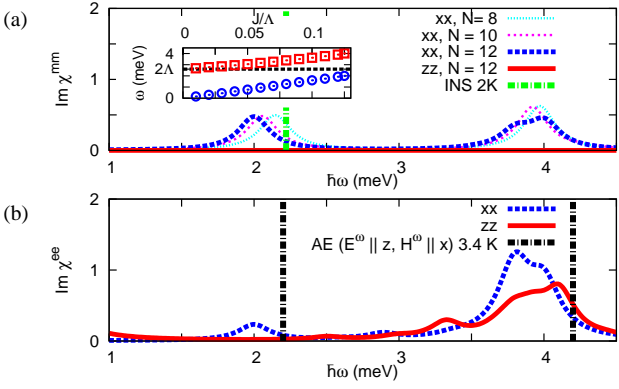}   
    \end{center}
    \caption{(Color online) 
      (a) Im $\chi^{\rm mm}_{\gamma\gamma} (\omega)$ for 
      $\gamma = x$, and $z$. 
      Magnon energy observed in INS
      is extracted from Ref.~\onlinecite{zheludev03}.
      Inset: Peak positions around $2$ and $4$ meV 
      for Im $\chi^{\rm mm}_{xx} (\omega)$
      as a function of $J/\Lambda$ on $12$-site cluster.	
      (b) Im $\chi^{\rm ee}_{\gamma\gamma} (\omega)$ for 
      $\gamma = x$, and $z$ on $12$-site cluster.
      Peak positions observed in THz AE for $E^\omega_z$ and $H^\omega_x$
      polarizations are extracted from Ref.~\onlinecite{kezsmarki10}.
    }
    \label{fig:epsilon_mu}
  \end{figure}

  To clarify the absorption processes at THz frequencies in detail,
  we investigate an $S=3/2$ Heisenberg model
  on a square lattice with the uniaxial anisotropy term
  under external magnetic field ${\bf B}^{\rm ex}$:
  \begin{eqnarray}
    {\cal H} & = & \sum_{n.n.} \left\{  J \,\, {\bf S}_i \cdot {\bf S}_j 
%    + \bf{D} \cdot ({\bf S}_i \times {\bf S}_{j}) \right\} \nonumber \\
    - D^z_{ij} \, (S^x_i S^y_{j} - S^y_i S^x_{j}) \right\} 
    \nonumber \\ & &
    + \sum \left\{ \Lambda (S_i^z)^2   
    - g \mu_B {\bf B}^{\rm ex} \cdot {\bf S}_i\right\},
    \label{eq:model}
  \end{eqnarray}
  where ${\bf S}_i$ is an $S = 3/2$ spin operator on $i$-site.
  The directions of Dzyaloshinsky-Moriya (DM) interactions $D^z_{ij}$ 
  on each bond can be determined uniquely from the crystal structure
  as in Fig.~\ref{fig:model} (a)~\cite{memo-DM}.
  The DM interaction lifts the two-fold degeneracy
  in antiferromagnetic ordered states.
  Reflecting the rotation of ${\rm CoO_4}$ tetrahedron around $z$ axis,
  local EPs on $i$-site are given by 
  $p_i^{x} = -K \,\,[ \cos (2 \kappa_i) ( S_i^z S_i^{x} + S_i^{x} S_i^{z} ) 
  + \sin (2 \kappa_i) (S_i^{y} S_i^{z} + S_i^z S_i^{y}) ]$,
  $p_i^{y} = K \,\,[ \cos (2 \kappa_i) (S_i^{y} S_i^{z} + S_i^z S_i^{y})
  + \sin (2 \kappa_i)      ( S_i^z S_i^{x} + S_i^{x} S_i^{z} )  ]$, and
  $p_i^{z}  = K \,\,[ \cos (2 \kappa_i) \{ (S_i^{y})^2 - (S_i^{x})^2 \} 
  - \sin (2 \kappa_i) (S_i^{x} S_i^{y} + S_i^{y} S_i^{x} ) ]$,
  where $\kappa_i$ is the rotation angle with
  $\kappa_i =  \kappa (-\kappa)$ on the $A$($B$)-sublattice 
  as in Fig.~\ref{fig:model} (a)~\cite{murakawa10}.
  Magnetization and EP are defined as
  $M^\gamma = \sum g \mu_B \mu_0 S^{\gamma}_i$ and
  $P^\gamma = \sum p_i^\gamma$ ($\gamma = x, y$, and $z$), respectively.
  ${\bf M}$ and ${\bf P}$ under the in-plane external magnetic field 
  ${\bf B}^{\rm ex} \| (\cos \phi_{\rm B}, \sin \phi_{\rm B}, 0)$
  have been calculated by an exact 
  diagonalization on $N$-site clusters ($N = 8, 10$, and $12$). 
  To reproduce $M^x$ and $P^z$ observed 
  in Refs.~\onlinecite{murakawa10} and \onlinecite{murakawa10b},  
  the parameters in Eq.(\ref{eq:model}) are estimated as
  $J/\Lambda = 0.125$, $|D^z|/\Lambda = 0.005$,
  $\Lambda = 1.3$ meV, $\kappa = \pi/8$,
  and $K = 3.8 \times 10^{-32}$ C$\cdot$m.
  Here we use $g =2$ while
  $V = 1.0 \times 10^{-28}$ ${\rm m}^3$ is the volume per Co.
  As typical examples,
  the magnetization curve along ${\bf B}^{\rm ex} \| x$ and
  magnetic field direction dependence of $P^z$
  at $B^{\rm ex} = 5$~T on a $12$-site cluster
  are shown in  Figs.~\ref{fig:model} (b) and (c), respectively.
  Here, system size effects are found to be negligibly small.
  Note that the magnon energy observed
  in the INS~\cite{zheludev03} at 2 K
  is also reproduced with this parameter set 
  (see Fig.~\ref{fig:epsilon_mu} (a)).
  We have confirmed that the results are qualitatively
  robust against choices of the parameters within the
  strong anisotropy limit $J/\Lambda \ll 1$.

  Let us consider excitation processes by
  magnetic components of EMW, i.e.,
  M1 transitions, which are related to the
  imaginary part of the magnetic susceptibility:
  \begin{eqnarray}
    {\rm Im} \,\chi^{\rm mm}_{\gamma\gamma} (\omega) & = & 
     \frac{\pi}{\hbar N V \mu_0}   \sum_n
    \left| \langle n | M^\gamma | 0 \rangle \right|^2
    \delta(\omega - \omega_{n0}). 
    \label{eq:mu} 
  \end{eqnarray}
  Here $| 0 \rangle$ is the ground state,
  $| n \rangle$ are excited states and 
  $\hbar \omega_{n0}$ are excitation energies to $| n \rangle$
  while $\gamma = x, y$, and $z$.
  Eq.~(\ref{eq:mu}) is calculated on $N$-site clusters ($N = 8, 10$, and $12$)
  by the Lanczos method~\cite{dagotto94},
  where the $\delta$-function is replaced by a Lorentzian 
  with a width $\epsilon/\Lambda = 0.1$.
  The results at $B^{\rm ex} = 0$
  are shown in Fig.~\ref{fig:epsilon_mu} (a).
  Out of plane component ${\rm Im} \,\chi^{\rm mm}_{zz} (\omega)$ vanishes.
  In-plane components are found to be identical, 
  ${\rm Im} \,\chi^{\rm mm}_{yy} (\omega) = {\rm Im} \,\chi^{\rm mm}_{xx} (\omega)$.
  They show that the magnetic components $H^{\omega}_{x}$ and $H^{\omega}_{y}$
  induce magnetic resonances at around 2 meV and 4 meV
  (Fig.~\ref{fig:epsilon_mu} (a)).
  As shown in the figure,
  the system size effects are small.
  Hereafter, we show the results on the $12$-site cluster. 
  We indeed see that
  the excitation around 2~meV corresponds to one of the spin-wave branches
  observed in the INS~\cite{zheludev03}, while the higher energy mode 
  is an excitation accompanied with the anisotropy gap 
  excitation $2 \Lambda$. 
  These features are clarified from 
  $J$ dependence of the peak positions.
  As shown in the inset of Fig.~\ref{fig:epsilon_mu} (a),
  in decreasing $J$, the high energy peak continuously shifts 
  to single site gap excitation $2\Lambda = 2.6$~meV,
  whereas the low energy peak position is proportional to~$J$.

  When spin states couple to electric fields through EP,
  E1 process may excite magnetic excitations~\cite{tanabe65,katsura07}.
  Such processes can be clarified from the dielectric susceptibility via
  spin-dependent EP
  \begin{eqnarray}
    {\rm Im} \, \chi^{\rm ee}_{\gamma\gamma} (\omega) & = & 
    \frac{\pi}{\hbar N V \epsilon_0}
    \sum_n \left| \langle n | P^\gamma | 0 \rangle \right|^2
    \delta(\omega - \omega_{n0}). 
    \label{eq:epsilon} 
  \end{eqnarray}
  At $B^{\rm ex} = 0$, in-plane components of dielectric susceptibility
  are found to be uniform, 
  ${\rm Im} \,\chi^{\rm ee}_{yy} (\omega) = {\rm Im} \,\chi^{\rm ee}_{xx} (\omega)$,
  as in the case for the magnetic susceptibility.
  Contributions to the 2~meV absorption are small.
  The 4~meV resonance is active for any electric components
  (see Fig.~\ref{fig:epsilon_mu} (b)).
  From these results,
  we conclude that the selection rules 
  and the peak positions are consistent with those obtained
  in the THz AE~\cite{kezsmarki10}.

  The temperature dependence of THz AE can also be explained
  qualitatively.
  In Ref.~\onlinecite{kezsmarki10}, 4 meV absorption is observed
  even above $T_N$, whereas
  absorption at 2~meV vanishes at $T_N$ upon increasing the temperature.
  The anisotropy gap excitation energy 
  $2 \Lambda \sim 30$ K is larger than $T_N$,
  and such a resonance can be observed even 
  above N\'{e}el temperature, i.e., $T_N < T \lesssim 2 \Lambda$.
  However, the resonance at 2~meV 
  vanishes above $T_N$, since the spin wave excitation 
  exists only in the ordered state.

\begin{figure}
    \begin{center}
      \includegraphics[width=0.9\columnwidth]{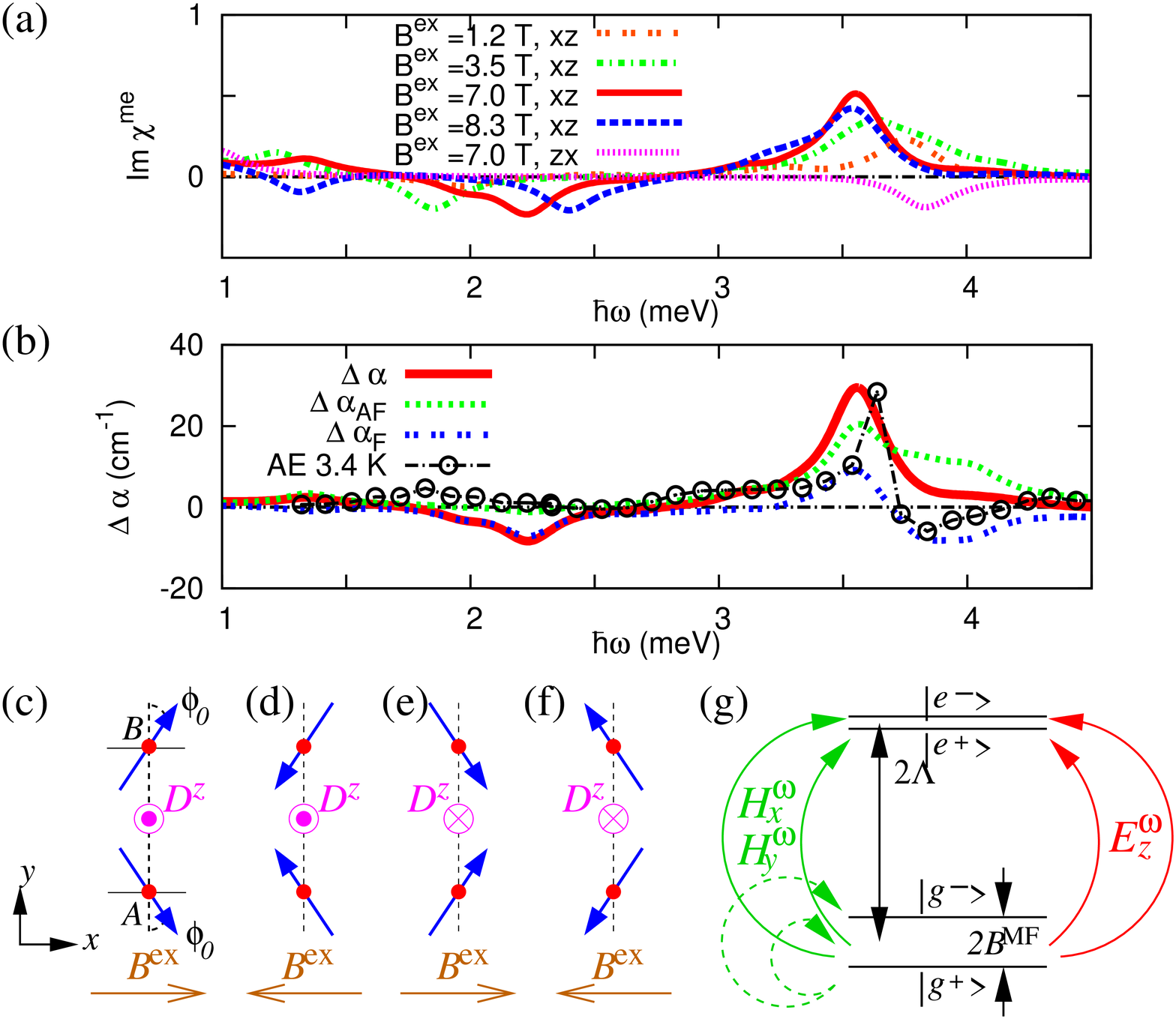}   
    \end{center}
    \caption{(Color online) 
      (a) Im $\chi^{\rm me}_{xz} (\omega)$ under
      the external magnetic fields ${\bf B}^{\rm ex} \| x$.
      At $7$ T, Im $\chi^{\rm me}_{zx} (\omega)$ is also shown.
      (b) Difference of the absorption coefficient
      by EMW propagating directions $\Delta \alpha (\omega)$ 
      at $B^{\rm ex}_x = 7$ T, which can be decomposed to 
      ferromagnetic component $\Delta \alpha_{\rm F}(\omega)$
      and antiferromagnetic one $\Delta \alpha_{\rm AF}(\omega)$. 
      $\Delta \alpha (\omega)$ obtained 
      from THz AE for $E^\omega_z$ and $H^\omega_x$
      are extracted from Ref.~\onlinecite{kezsmarki10}.
      (c)-(f) Two sublattice spin structures in $B^{\rm ex}_x$. 
      See text for detail.
      (g) Excitation process in the limit  $B^{\rm MF}/\Lambda \ll 1$.
    }
    \label{fig:alpha}
  \end{figure}

  In practice,  M1 and E1 processes are invoked 
  through the interaction with EMW as
  ${\cal H}^\prime = - {\bf E}^\omega \cdot {\bf P} 
  - {\bf H}^\omega \cdot {\bf M}$, where
  ${\bf E}^\omega$ (${\bf H}^\omega$) is the
  electric (magnetic) component of EMW.
  Provided that both M1 and E1 processes induce an identical excitation,
  there is a cross correlation between magnetic and 
  electric components of EMW, 
  i.e., the interference between electric and magnetic responses.
  As we show details in the following, the effects of the interference 
  can be observed directly as the linear NDD, e.g.,
  the interference enhances absorption intensity 
  for the EMW with a propagation vector $+{\bf k}$
  but weakens that for the EMW with $-{\bf k}$,
  since reversing ${\bf k}$ is equivalent 
  to reversing the relative sign of ${\bf E}^\omega$ and ${\bf H}^\omega$
  due to ${\bf H}^\omega = (1/\mu_0 \omega) {\bf k} \times {\bf E}^\omega$.
  As a typical case, we consider 
  dynamical magnetoelectric susceptibility
  for $M^x$ and $P^z$
  \begin{eqnarray}
    {\rm Im} \chi^{\rm me}_{x z} (\omega) 
    & = & \sum_n  
    \frac{\pi c}{2 \hbar N V}
    \left(\!\frac{}{} \langle 0 | M^x | n  \rangle \langle n | P^z  | 0 \rangle
    \right.
    \nonumber \\     
    && \hspace*{0.5cm}\left. + \langle 0 | P^z | n  \rangle 
    \langle n | M^x  | 0 \rangle \frac{}{} \!
    \right) \delta(\omega - \omega_{n0}),
    \label{eq:chi_me}
  \end{eqnarray}   
  where $c \equiv 1/\sqrt{\epsilon_0 \mu_0}$.
  The results under ${\bf B}^{\rm ex} \| x$ are
  shown in Fig.~\ref{fig:alpha} (a). 
  ${\rm Im} \chi^{\rm me}_{x z} (\omega)$ is enhanced around 4 meV excitation.
  Note that ${\rm Im} \chi^{\rm me}_{zx} (\omega)$ is 
  much smaller than ${\rm Im} \chi^{\rm me}_{xz} (\omega)$. 
  ${\rm Im} \chi^{\rm me}_{zx} (\omega)$ at 7 T is also shown 
  in Fig.~\ref{fig:alpha} (a). 

  Experimentally, such a cross correlated effect
  can be observed as the linear NDD~\cite{kezsmarki10}.  
  By introducing a complex refractive index $N$,
  a polarized plane wave with ${\bf E}^{\omega} \| z$, ${\bf H}^{\omega} \| x $ and
  ${\bf k} \| y$ is described as 
  $E_z^\omega = E_0^z \exp(-i \omega ( t - (N y /c) ))$ and
  $H_x^\omega = H_0^x \exp(-i \omega ( t - (N y /c) ))$.
  From the Maxwell's equations,
  $N$ is given as
  $N^{\pm} \sim \sqrt{\epsilon_{zz} (\omega) \mu_{xx}(\omega)}
  \pm \chi^{\rm me}_{x z} (\omega)$ 
  where  $\mu_{xx} (\omega) = 1 + \chi^{\rm mm}_{xx} (\omega)$ and
  $\epsilon_{zz} (\omega) = 1 + \chi^{\rm ee}_{zz} (\omega)$.
  Here, $N^{+}$ ($N^{-}$) is a complex refractive index for EMW
  propagating to $+y$ ($-y$) direction.
  Thus, non-reciprocal part of an absorption coefficient
  $\alpha^{\pm}(\omega) = (\omega/c) \, {\rm Im} N^{\pm}$
  is given by
  $\Delta \alpha(\omega) = \alpha^{+}(\omega) - \alpha^{-}(\omega) =
  (2 \omega/c) \, {\rm Im} \chi^{\rm me}_{x z} (\omega)$.
  $\Delta \alpha(\omega)$ 
%  at $B^{\rm ex}_{x} = 7$ T is shown in Fig.~\ref{fig:alpha} (b)
  at $B^{\rm ex}_{x} = 7$ T is shown in Fig.~\ref{fig:alpha} (b),
  where the peak position and the magnitude of $\Delta \alpha (\omega)$
  are consistent with those observed in THz AE~\cite{kezsmarki10}.
%  and the peak position is consistent with 
%  that observed in THz AE~\cite{kezsmarki10}.
%  The discrepancy of the magnitude of $\Delta \alpha (\omega)$
%  may be related to the
%  temperature effects: the calculation is performed at $T = 0$,
%  whereas the measurement 
%  at rather high temperature $T = 3.4$ K ($\sim 0.5 T_N$)~\cite{kezsmarki10}.

  Let us consider the magnetic origin for non-reciprocal part
  of the absorption coefficient 
  $\Delta \alpha (\omega) 
  \sim (2 \omega/c) \, {\rm Im} \chi^{\rm me}_{x z} (\omega)$.
  Under the external magnetic field ${\bf B}^{\rm ex} \perp z$, 
  spin structure in the N\'{e}el ordered state is
  uniquely determined due to an energy gain of the DM term,
  e.g., the state in Fig.~\ref{fig:alpha} (c)
  is stabilized by ${\bf B}^{\rm ex} \| x$ with $B^{\rm ex}_x > 0$.
  The EMW propagating to $-y$ direction 
  in Fig.~\ref{fig:alpha} (c)
  corresponds to that propagating to $+y$ direction 
  in Fig.~\ref{fig:alpha} (d) which is realized by
  a $180^\circ$ rotation around $z$ axis on a spin site.
  Thus, reversing the magnetic field
  $B^{\rm ex}_x \rightarrow -B^{\rm ex}_x$ is equivalent to 
  reversing the EMW direction $k^y \rightarrow -k^y$,
  which is consistent with the experimental 
  observation~\cite{kezsmarki10}.
  Note that, when we reverse the magnetic field 
  $B^{\rm ex}_x \rightarrow -B^{\rm ex}_x$,
  both ferromagnetic moment $m^x_u \equiv S^x_A + S^x_B$ 
  and antiferromagnetic component $m^y_{s} \equiv S^y_A - S^y_B$ 
  change their sign as shown in Figs.~\ref{fig:alpha} (c) and (d). 
  Each contribution to $\Delta \alpha (\omega)$ 
  can be obtained by changing the sign of $D^z$
  in the calculation, since only $m^y_s$ ($m^x_u$) changes its sign
  between states in Figs.~\ref{fig:alpha} (c) and (e) ((c) and (f)). 
  As a result, $\Delta \alpha (\omega)$ is decomposed into two parts
  $\Delta \alpha_{\rm F}(\omega)$ and $\Delta \alpha_{\rm AF}(\omega)$, which
  depend on the modification of $m^x_u$ and $m^y_s$, respectively.
  By comparing absorptions for the states 
  in Figs~\ref{fig:alpha} (c)-(f),
  $\Delta \alpha_{F}(\omega)$ and $\Delta \alpha_{\rm AF}(\omega)$ are extracted as
  in Fig.~\ref{fig:alpha} (b).
  The results indicate that $\Delta \alpha_{\rm AF}(\omega)$
  is dominant for the NDD around 4~meV.
  Generally, NDD can be realized when spontaneous
  magnetization and EP coexist.
  In the present model, 
  however, realization of a N\'{e}el ordered state
  without ferromagnetic moment is sufficient to induce NDD. 
  Once a single domain structure of the N\'{e}el ordered state
  is realized,  $\Delta \alpha(\omega)$ can be finite even 
  at $B^{\rm ex}_x \rightarrow 0$ 
  and $|D^z_{ij}| \rightarrow 0$.

  Finally, we note that the selection rules and
  the cross correlated effects
  can qualitatively be determined within a mean field (MF) approximation.
  The spin Hamiltonian (\ref{eq:model})
  can be approximated as
  ${\cal H}^{\rm MF} = \sum \{ \Lambda (S_i^{z})^2 
  - g \mu_B {\bf B}^{\rm MF}_i \cdot {\bf S}_i\}$,
  where $g \mu_B {\bf B}^{\rm MF}_i \equiv  
  g \mu_B {\bf B}^{\rm ex} - \sum_j \{J \langle {\bf S}_j \rangle
  \mp |D_{ij}^z| (\langle S_j^y \rangle, -\langle S_j^x \rangle, 0)\}$
  ($-$($+$) for the $i$-site on $A$($B$)-sublattice).
  For simplicity, spin states
  under ${\bf B}^{\rm MF}_i = (B^{\rm MF} \cos \phi_i, B^{\rm MF} \sin \phi_i, 0)$
  in the limit $B^{\rm MF}/\Lambda \ll 1$ are discussed.
  Four eigenstates 
  at site $i$ are given in a form:
  $|g_i^\pm \rangle  =  \frac{1}{\sqrt{2}} (
  e^{-i \phi_i/2} |\frac{1}{2} \rangle \pm 
  e^{i \phi_i/2} |-\frac{1}{2} \rangle )$ and 
  $|e_i^\pm \rangle = \frac{1}{\sqrt{2}} (
  e^{-i 3 \phi_i/2} |\frac{3}{2} \rangle \pm 
  e^{i 3 \phi_i/2} |-\frac{3}{2} \rangle )$
%  \begin{eqnarray}
%    |g_i^\pm \rangle & = & \frac{1}{\sqrt{2}} \left(
%    e^{-i \frac{\phi_M}{2}} |\frac{1}{2} \rangle \pm 
%    e^{i \frac{\phi_M}{2}}|-\frac{1}{2} \rangle \right),  \\
%    |e_i^\pm \rangle & = & \frac{1}{\sqrt{2}} \left(
%    e^{-i \frac{3 \phi_M}{2}} |\frac{3}{2} \rangle \pm 
%    e^{i \frac{3 \phi_M}{2}} |-\frac{3}{2} \rangle \right),
%  \end{eqnarray}
  as in Fig.~\ref{fig:alpha} (g).
  Eigenenergies for $|g_i^\pm \rangle$ are
  $\Lambda/4 \mp g \mu_B B^{\rm MF}$ and
  for both $|e_i^\pm \rangle$, $9 \Lambda/4$. 
  As a typical example, 
  let us consider excitation processes induced by
  $H_x^\omega$, $H_y^\omega$ and $E_z^\omega$.
  From the ground state $|g_i^+ \rangle$,
  processes through $S^{x}_i$, $S^{y}_i$ and $p^{z}_i$ are
  \begin{eqnarray}
    S^{x}_i |g_i^+ \rangle & = &  \cos\phi_i |g_i^+ \rangle 
    + i \sin \phi_i |g_i^- \rangle \nonumber \\
    && + \sqrt{3} ( \cos\phi_i |e_i^+ \rangle 
    + i \sin\phi_i |e_i^- \rangle )/2,
    \label{eq:sx-single} \\
    S^{y}_i |g_i^+ \rangle & = &  \sin\phi_i |g_i^+ \rangle 
    - i \cos \phi_i |g_i^- \rangle \nonumber \\
    && + \sqrt{3} ( \sin\phi_i |e_i^+ \rangle 
    - i \cos\phi_i |e_i^- \rangle )/2, 
    \label{eq:sy-single} \\
    p^{z}_i |g_i^+ \rangle & = & 
    - \sqrt{3} K \cos (2 \phi_i - 2 \kappa_i) |e_i^+ \rangle 
    \nonumber \\  && 
    - i \sqrt{3} K \sin (2 \phi_i - 2 \kappa_i) |e_i^-  \rangle. 
%    p^{z}_i |g_i^+ \rangle & = & 
%    - \sqrt{3} K (\cos 2 (\phi_i - \kappa_i) |e_i^+ \rangle
%    + i \sin 2 (\phi_i - \kappa_i) |e_i^-  \rangle).  
    \label{eq:e-single}
  \end{eqnarray}
  We see that $E^{\omega}_z$ 
  can only induce magnetic excitations
  with the anisotropy gap $2 \Lambda$.
  This is consistent with the results in Fig.~\ref{fig:epsilon_mu} (b).
  For the EMW with ${\bf E}^\omega \| z$ and
  ${\bf H}^\omega \| (\cos \phi_\omega, \sin \phi_\omega, 0)$,
  the cross correlation effect is qualified by
  a spectral weight $\Delta I(\phi^\omega)  
% = \int (\Delta\alpha(\omega)/\omega) d \omega
  \propto \int (\cos \phi_\omega {\rm Im} \chi^{\rm me}_{x z} (\omega)   
  +  \sin \phi_\omega {\rm Im} \chi^{\rm me}_{y z} (\omega)  ) d \omega$.
  The canted N\'{e}el ordered state under
  ${\bf B}^{\rm ex} \| (\cos\phi_{\rm B}, \sin\phi_{\rm B}, 0)$
  gives $\phi_i = \phi_{\rm B} \mp (\pi/2 - \phi_0)$ 
  ($-$ ($+$) for the spin on $A$($B$)-sublattice),
  where $\phi_0$ is a spin canting angle.
  By applying Eqs.~(\ref{eq:sx-single})-(\ref{eq:e-single}) to
%  $H^{\omega x}$ and $H^{\omega y}$ induced excitations are
  Eq.~(\ref{eq:chi_me}), we obtain that
  \begin{equation}
    \Delta I (\phi_\omega) \propto \cos(\phi_\omega + \phi_{\rm B}) 
    \sin (2 \kappa - \phi_0).
  \end{equation}
  For $\phi_{\rm B} = 0$ (a state in Fig.~\ref{fig:alpha} (c)),
  we obtain that $ \Delta I (0) \propto \sin(\phi_0 - 2 \kappa)$ 
  (for ${\bf H}^\omega \| x$) and
  $ \Delta I (\pi/2) = 0$ (for ${\bf H}^\omega \| y$)
  as already expected from the 
  symmetry argument in Ref.~\onlinecite{kezsmarki10}.   
  We see $\Delta I (0) \neq 0$ even for $\phi_0 = 0$, which 
  indicates the existence of the NDD in a collinear N\'{e}el ordered state.

  In addition, we can predict 
  the NDD for ${\bf k} \| [010]$,
  ${\bf E}^\omega \| [001]$, and  ${\bf H}^\omega \| [100]$ ($\phi_\omega = -\pi/4$) 
  under ${\bf B}^{\rm ex} \| [010]$ ($\phi_{\rm B} = \pi/4$), 
  although there is no spontaneous EP~\cite{murakawa10} for 
  this ${\bf B}^{\rm ex}$ direction.
  In fact, $\chi^{\rm me}_{[100][001]}(\omega)$ under  ${\bf B}^{\rm ex} \| [010]$
  is found to be non-zero in the numerical calculation. 
  Observation of the NDD in this 
  condition is a crucial test for the validity of our theory.
  As another example,
  we can easily derive
  $\Delta I_{zx} \propto \int {\rm Im} \chi^{\rm me}_{z x} (\omega) d \omega = 0$
  for the EMW  with ${\bf E}^\omega \| x$ and ${\bf H}^\omega \| z$ under ${\bf B}^{ex} \| x$,
  which is consistent with the results calculated
  at $B^{\rm ex}_x = 7$ T:
  $\int {\rm Im} \chi^{\rm me}_{zx}(\omega) d \omega \ll
  \int {\rm Im} \chi^{\rm me}_{xz}(\omega) d \omega$ 
  (see Fig.~\ref{fig:alpha}~(a)).

  Our results indicate the potential of 
  the spin-dependent metal-ligand hybridization mechanism for 
  novel absorption processes which
  might be observed in a wide range of materials
  with a spin at a site without inversion center, e.g.,
  in a tetrahedron and a pyramid of ligand atoms.
%  Similar magnetoelectric effects 
%  observed in ${\rm Ca_2CoSi_2O_7}$~\cite{akaki09}
%  can be explained by the concepts in this Letter.

  We thank I. K\'{e}zsm\'arki, N. Kida, S. Bord\'acs
  H. Murakawa, Y. Onose, T. Arima, R. Shimano,
  and Y. Tokura for fruitful discussion.
  This work is in part supported by Grant-In-Aids for Scientific 
  Research from the Ministry of Education, Culture, 
  Sports, Science and Technology (MEXT) Japan.
  
  \bibliographystyle{apsrev}
%  \bibliography{EM}
  \bibliography{EM_short}

\begin{thebibliography}{25}
\bibitem[{mem({\natexlab{a}})}]{memo-Pz}
\bibinfo{note}{$p^z$ is suppressed as $B^{\rm ex}_x$ is decreased below $\sim
  1$ T and deviates from the theoretical estimates based on a classical spin
  picture~\cite{murakawa10}. However, the suppression of $p^z$ at $B^{\rm ex}_x
  = 0$ can be understood by considering the effects of quantum fluctuation in a
  limit $\Lambda \rightarrow \infty$ on a single site, since $\langle p^z
  \rangle$ vanishes for any linear combinations of $|\pm \frac{1}{2} \rangle$.}
\bibitem[{mem({\natexlab{b}})}]{memo-DM}
\bibinfo{note}{$D^x$ ($D^y$) component on the bond along $y$ ($x$) direction is
  also non-zero, but the effects of them are found to be negligible in the
  calculations.}
\end{thebibliography}

\end{document}